\documentclass[aps,showpacs,nofootinbib,prd,twocolumn]{revtex4-1}
\usepackage{graphicx}
\usepackage{amssymb}
\usepackage{amsmath}
\usepackage{xcolor}
\usepackage{mathtools,slashed}
\usepackage{epstopdf}
\usepackage[utf8]{inputenc}
\usepackage{url}
\usepackage{subfigure}
\usepackage{soul}
\usepackage{float}
\usepackage{newtxmath}

\usepackage{epsfig}

\newcommand{\be}{\begin{eqnarray}}
\newcommand{\ee}{\end{eqnarray}}
\begin{document}

\title{ Plasma screening and the critical end point in the QCD phase diagram}


\author{Alejandro Ayala$^{1,2}$}
\author{Bilgai Almeida Zamora$^3$}
\author{J. J. Cobos-Mart\'inez$^4$}
\author{S. Hern\'andez-Ortiz$^5$}
\author{L. A. Hernández$^{6,2,7}$}
\author{Alfredo Raya$^{8,9}$}
\author{Mar\'ia Elena Tejeda-Yeomans$^{10}$}
  \address{
  $^1$Instituto de Ciencias
  Nucleares, Universidad Nacional Aut\'onoma de M\'exico, Apartado
  Postal 70-543, CdMx 04510,
  Mexico.\\
  $^2$Centre for Theoretical and Mathematical Physics, and Department of Physics,
  University of Cape Town, Rondebosch 7700, South Africa.\\
  $^3$Departamento de Investigaci\'on en F\'isica, Universidad de Sonora, Boulevard Luis Encinas J. y Rosales, 83000, Hermosillo, Sonora, Mexico.\\
  $^4$Departamento de F\'isica, Universidad de Sonora, Boulevard Luis Encinas J. y Rosales, 83000, Hermosillo, Sonora, Mexico.\\
  $^5$Institute for Nuclear Theory, University of Washington, Seattle, WA, 98195, USA.\\
  $^6$Departamento de F\'isica, Universidad Aut\'onoma Metropolitana-Iztapalapa, Av. San Rafael Atlixco 186, C.P, CdMx 09340, Mexico.\\
  $^7$ Facultad de Ciencias de la Educaci\'on, Universidad Aut\'onoma de Tlaxcala, Tlaxcala, 90000, Mexico. \\
  $^8$Instituto de F\'isica y Matem\'aticas, Universidad Michoacana de San Nicol\'as de Hidalgo, Edificio C-3, Ciudad Universitaria, Francisco J. M\'ujica s/n Col. Fel\'icitas del R\'io. C. P. 58040, Morelia, Michoac\'an, Mexico.\\
  $^9$Centro de Ciencias Exactas, Universidad del Bío-Bío. Avda. Andrés Bello 720, Casilla 447, 3800708, Chillán, Chile.\\
  $^{10}$Facultad de Ciencias - CUICBAS, Universidad de Colima, Bernal D\'iaz del Castillo No. 340, Col. Villas San Sebasti\'an, 28045 Colima, Mexico.
  }
\begin{abstract}
In heavy-ion collisions, fluctuations of conserved charges are known to be sensitive observables to probe criticality for the QCD phase transition and to locate the position of the putative critical end point (CEP). In this work we seek to show that the Linear Sigma Model with quarks produces an effective description of the QCD phase diagram in which deviations from a  Hadron Resonance Gas are due to plasma screening effects, encoded in the contribution of the ring diagrams. Accounting for these, it is possible to include in the description the effect of long-range correlations. To set the model parameters we use LQCD results for the crossover transition at vanishing chemical potential. Finally, studying baryon number fluctuations from the model, we show that  the CEP can be located within the HADES and/or the lowest end of the NICA energy domain, $\sqrt{s_{NN}}\sim 2$ GeV.
\end{abstract}
\maketitle

\section{Introduction}
In recent years, the study of hadron matter under extreme conditions of temperature and baryon density has become a subject of great interest. Of particular importance is the possibility of experimentally exploring the QCD phase structure by means of relativistic heavy-ion collisions. Currently, this exploration is carried out in experimental facilities such as RHIC, with the STAR Beam Energy Scan program, and HADES~\cite{STAR:2020tga,HADES:2020wpc}. Dedicated experiments soon to come on line, such as the NICA-MPD~\cite{Kekelidze:2017tgp,MPD:2022qhn} and FAIR-CBM~\cite{Senger:2017nvf}, are expected to widen the energy range for the exploration of the phase diagram.

From Lattice QCD (LQCD) calculations, its known that for finite temperature $T$ and vanishing baryon chemical potential $\mu_B$, the transition between the confined/chiral symmetry broken and the deconfined/partially restored chirally symmetric phases, is a crossover that occurs  at a pseudocritical temperature $T_c(\mu_B=0)\simeq 158$ MeV~\cite{Borsanyi:2020fev,Bazavov:2018mes,Aoki:2006we}. Also from the calculations made using effective models~\cite{Roessner:2006xn,Ayala:2019skg, Asakawa:1989bq,Ayala:2017ucc,Gao:2016qkh,Gao:2020fbl}, it can be found that this transition becomes first order at low $T$ and high $\mu_B$. Therefore, the point at which the first-order phase transition line in the $T$ vs $\mu_B$ plane ends and the crossover begins, as the temperature increases, is called the Critical End Point (CEP). Unfortunately, LQCD calculations cannot be used to directly determine the position of this CEP due to the severe sign problem~\cite{Ding:2015ona} but results employing the Taylor series expansion around $\mu_B=0$ or the extrapolation from imaginary to real $\mu_B$ values, suggest that the CEP has not yet been found for $\mu_B/T\leq 2$ and $145\leq T\leq 155$ MeV~\cite{Sharma:2017jwb,Borsanyi:2020fev,Borsanyi:2021sxv}.

The Hadron Resonance Gas Model (HRGM) describes the crossover transition line for low values of $\mu_B$ found by LQCD~\cite{Karsch:2016yzt} when the occupation numbers are given in terms of Boltzman statistics~\cite{Braun-Munzinger:2003pwq}. Therefore, the strategy to locate the CEP consists in finding deviations from the statistical behavior of the HRGM. The statistical properties of a thermal system are characterized in terms of the cumulants of its conserved charges, that are extensive quantities~\cite{Asakawa:2015ybt}. To avoid uncertainties introduced by volume effects, the analyses involve ratios of these cumulants. The strategy narrows down to find deviations in the ratios of these cumulants from those obtained in HRGM, which are described by the Skellam distribution where the ratios of cumulants of even order are equal to 1. The baryon number is a conserved quantity that can be experimentally probed by means of measurements of proton multiplicities~\cite{STAR:2020ddh,STAR:2021iop,STAR:2021rls,HADES:2020wpc}. Therefore the location of the CEP can be identified by the appearance of critical behavior~\cite{Hatta:2002sj, Stephanov:1999zu,Bzdak:2016sxg,Bzdak:2019pkr,Athanasiou:2010kw,Mroczek:2020rpm} in this and other conserved charges such as electric charge and strangeness, when a collision energy scan is performed. 

In this work we study baryon number fluctuations as a function of the collision energy looking at the evolution of cumulant ratios, in the context of relativistic heavy ion collisions, using the Linear Sigma Model with quarks (LSMq) including the plasma screening effects. The latter are encoded in the ring diagram contribution to the effective potential which becomes a function of the order parameter after spontaneous chiral symmetry breaking. Therefore, the statistical properties of the system can be formulated in terms of fluctuations of this order parameter~\cite{Stephanov:2008qz,Stephanov:2011pb} when the collision energy $\sqrt{s_{NN}}$ and thus $T$ and $\mu_B$, are varied. To provide analytical insight, we employ the high temperature approximation and work in the chiral limit. Although these approximations have limitations in terms of the accuracy for the CEP localization, they are a useful guide for future more precise studies. The remaining of this work is organized as follows: In Sec.~\ref{secII} we describe the LSMq and compute the effective potential up to the ring diagrams contribution, which requires computation of the self-energies corresponding to the meson degrees of freedom. The parameters of the model are fixed by requiring that the phase transition line in the vicinity of $\mu_B=0$ corresponds to the one found by the most recent LQCD calculations~\cite{Borsanyi:2020fev}. In Sec.~\ref{secIII}, we formulate how baryon number fluctuations are described in terms of the probability distribution associated with the order parameter near the transition line and report the results obtained from the analysis of the cumulant ratios used to describe baryon number fluctuations as a function of $\sqrt{s_{NN}}$. We show that these ratios deviate from the expectations of the HRGM for energies around $\sqrt{s_{NN}}\sim 4-6$ GeV and that the CEP can be found at energies $\sqrt{s_{NN}}\sim 2$ GeV. The model thus predicts that the CEP can be found either in the lowest NICA or within the HADES energy domain. We finally summarize in Sec.~\ref{concl}. The complete analysis can be found in Ref.~\cite{Ayala2022}. 

\section{Linear Sigma Model with quarks}~\label{secII}

The QCD phase diagram can be partially described by effective models; they can be used to explore different regions in parameter space depending on the degrees of freedom in the model. Given that LQCD calculations find that coincident deconfinement and chiral symmetry restoration transitions lines, it should be possible to explore the phase diagram emphasizing the chiral aspects of the transition only, like LSMq does. The Lagrangian of the LSMq is given by
\begin{eqnarray}
  \mathcal{L}&=&\frac{1}{2}(\partial_\mu \sigma)^2  + \frac{1}{2}(\partial_\mu \vec{\pi})^2 + \frac{a^2}{2} (\sigma^2 + \vec{\pi}^2)\nonumber\\
   &&- \frac{\lambda}{4} (\sigma^2 + \vec{\pi}^2)^2 + i \bar{\psi} \gamma^\mu \partial_\mu\psi -g\bar{\psi} (\sigma + i \gamma_5 \vec{\tau} \cdot \vec{\pi} )\psi ,\nonumber\\
\label{lagrangian}
\end{eqnarray}
where $\psi$ is a $SU(2)$ isospin doublet of quarks,
\begin{eqnarray}
 \vec{\pi}=(\pi_1, \pi_2, \pi_3 ),
\end{eqnarray} and $\sigma$ are isospin triplet and singlet,  corresponding to the three pions and the sigma meson, respectively. The squared mass parameter $a^2$ and the self-coupling $\lambda$ and $g$ are taken to be positive and, for the purpose of describing the chiral phase transition at finite $T$ and $\mu_B$, they need to be determined from conditions close to the phase boundary, and not from vacuum conditions. 
In order to allow for a spontaneous symmetry breaking, we work in the strict chiral limit and we let the $\sigma$ field to develop a vacuum expectation value $v$, namely,
\begin{eqnarray} \sigma \rightarrow \sigma + v,
\label{shift}
\end{eqnarray}
which can later be taken as the order parameter of the transition. After this shift, the Lagrangian can be rewritten as
\begin{eqnarray}
{\mathcal{L}}&=&\frac{1}{2}(\partial_\mu \sigma)^2  + \frac{1}{2}(\partial_\mu \vec{\pi})^2-\frac{1}
   {2}\left(3\lambda v^{2}-a^{2} \right)\sigma^{2}\nonumber\\
   &-&\frac{1}{2}\left(\lambda v^{2}- a^2 \right)\vec{\pi}^{2}+\frac{a^{2}}{2}v^{2} -\frac{\lambda}{4}v^{4}\nonumber\\
   &+& i \bar{\psi} \gamma^\mu \partial_\mu\psi
  -gv \bar{\psi}\psi + {\mathcal{L}}_{I}^b + {\mathcal{L}}_{I}^f,
  \label{lagranreal}
\end{eqnarray}
where ${\mathcal{L}}_{I}^b$ and  ${\mathcal{L}}_{I}^f$ are given by
\begin{align}
  {\mathcal{L}}_{I}^b&=-\frac{\lambda}{4}
  \sigma^4-\lambda v \sigma^3-\lambda v^3\sigma-\lambda\sigma^2 \pi^+\pi^--2\lambda v \sigma \pi^+\pi^- \nonumber \\
  &-\frac{\lambda}{2}\sigma^2(\pi^0)^2-\lambda v\sigma(\pi^0)^2-\lambda(\pi^+)^2(\pi^-)^2\nonumber \\
  &-\lambda \pi^+ \pi^-(\pi^0)^2-\frac{\lambda}{4}(\pi^0)^4+a^2v\sigma\nonumber\\
  {\mathcal{L}}_{I}^f&=-g\bar{\psi} (\sigma + i \gamma_5 \vec{\tau} \cdot \vec{\pi} )\psi.
  \label{lagranint}
\end{align}
The expressions in Eq.~(\ref{lagranint}) describe the interactions among the fields $\sigma$, $\vec{\pi}$ and $\psi$, after symmetry breaking.

From Eq.~(\ref{lagranreal}) we can see that the sigma, the three pions and the quarks have masses given by
\begin{align}
  m^{2}_{\sigma}&=3  \lambda v^{2}-a^{2},\ \ \ \ \ \
  m^{2}_{\pi}=\lambda v^{2}-a^{2}, \nonumber\\
  m_{f}&= gv,
\label{masses}
\end{align}
respectively. We study the behavior of the effective potential in order to analyze the chiral symmetry restoration conditions in terms of temperature and
quark chemical potential. The effective potential includes the classical
potential or tree-level contribution, the one-loop correction both for bosons and fermions and the ring diagrams contribution, which accounts for the plasma screening effects.

The tree-level potential is given by
\begin{equation}
    V^{\text{tree}}(v)=-\frac{a^2}{2}v^2+\frac{\lambda}{4}v^4,
    \label{treelevel}
\end{equation}
whose minimum is found at
\begin{equation}
    v_0=\sqrt{\frac{a^2}{\lambda}}.
\end{equation}
Since $v_0\neq 0$ then chiral symmetry is spontaneously broken. To include quantum corrections at finite temperature and density, we work within the imaginary-time formalism of thermal field theory. The general expression for the one-loop boson contribution can be written as
\begin{equation}
    V^{\text{b}}(v,T)=T\sum_n\int\frac{d^3k}{(2\pi)^3} \ln D_{\text{b}}(\omega_n,\vec{k})^{1/2},
    \label{1loopboson}
\end{equation}
where
\begin{equation}
D_\text{b}(\omega_n,\vec{k})=\frac{1}{\omega_n^2+k^2+m_b^2},
\end{equation} 
is the free boson propagator with $m_b^2$ being the square of the boson mass and $\omega_n=2n\pi T$ the Matsubara frequencies for boson fields.

For a fermion field with mass $m_f$, the general expression for the one-loop correction at finite temperature and quark chemical potential $\mu=\mu_B/3$ is
\begin{equation}
    V^{\text{f}}(v,T,\mu)=-T\sum_n\int\frac{d^3k}{(2\pi)^3} \text{Tr}[\ln S_\text{f}(\tilde{\omega}_n,\vec{k})^{-1}],
    \label{1loopfermion}
\end{equation}
where
\begin{equation}
S_\text{f}(\tilde{\omega}_n,\vec{k})=\frac{1}{\gamma_0( \tilde{\omega}_n+i\mu)+\slashed{k}+m_f},
\end{equation}
is the free fermion propagator and $\tilde{\omega}_n=(2n+1)\pi T$ are the Matsubara frequencies for fermion fields.
The ring diagrams term is given by
\begin{eqnarray}
    V^{\text{Ring}}(v,T,\mu)=\frac{T}{2}\sum_n\int\frac{d^3k}{(2\pi)^3}\ln [1+\Pi_{\text{b}}D_b(\omega_n,\vec{k})],\nonumber\\
    \label{rings}
\end{eqnarray}
where $\Pi_\text{b}$ is the boson self-energy. The self-energies for the sigma and pion fields are in general different. However, we are working in the high-temperature approximation, keeping only the leading matter effects. In this approximation the boson self-energies become mass independent and therefore independent of the boson species. Additionally, because no external agent that might couple to the electric charge, such as a magnetic field, is considered, there is no distinction between neutral and charged pions and we can write~\cite{Ayala:2021nhx}
\begin{eqnarray}
\Pi_{\text{b}}&\equiv&\Pi_\sigma=\Pi_{\pi^\pm}=\Pi_{\pi^0}\nonumber\\
&=&\lambda\frac{T^2}{2}-N_\text{f}N_\text{c}g^2\frac{T^2}{\pi^2}\left[\text{Li}_2\left(-e^{-\frac{\mu}{T}}\right)+\text{Li}_2\left(-e^{\frac{\mu}{T}}\right)\right],\nonumber\\
\label{Pileading}
\end{eqnarray}
where $N_\text{c}$ and $N_\text{f}$ are the number of colors and flavors, respectively, while ${\rm Li}_2(x)$ stands for the polylogarithm function of order 2.

Wrapping all the ingredients together, the effective potential at finite temperature and baryon density, up to the contribution of the ring diagrams, after renormalization of the boson and fermion masses in the $\overline{\text{MS}}$ scheme at the ultraviolet scale $\tilde{\mu}$, the effective potential can be written as
\begin{widetext}
\begin{align}
V^{\text{eff}}(v)&=-\frac{a^{2}}{2}v^{2}+\frac{\lambda}{4}v^{4}+\sum_{\text{b}=\pi^\pm,\pi^0,\sigma}\left\{-\frac{T^4\pi^2}{90}+\frac{T^2m_\text{b}^2}{24}-\frac{T(m_\text{b}^2+\Pi_\text{b})^{3/2}}{12\pi}-\frac{m_\text{b}^4}{64\pi^2}\left[\ln\left(\frac{\tilde{\mu}^2}{(4\pi T)^2}\right)+2\gamma_E\right]\right\}\nonumber\\
&+N_\text{c}N_\text{f}\left\{\frac{m_f^4}{16\pi^2}\Big[\ln\left(\frac{\tilde{\mu}^2}{T^2}\right)- \psi^{0}\left(\frac{1}{2}+\frac{i\mu}{2\pi T}\right)- \psi^{0}\left(\frac{1}{2}-\frac{i\mu}{2\pi T}\right)\right.+
\psi^{0}\left(\frac{3}{2}\right)-2\left(1+\ln(2\pi)\right)+\gamma_E\Big]\nonumber\\
&-\left.\frac{m_\text{f}^2T^2}{2\pi^2}\left[\text{Li}_2\left(-e^{-\frac{\mu}{T}}\right)+\text{Li}_2\left(-e^{\frac{\mu}{T}}\right)\right]+\frac{T^4}{\pi^2}\left[\text{Li}_4\left(-e^{-\frac{\mu}{T}}\right)+\text{Li}_4\left(-e^{\frac{\mu}{T}}\right)\right]\right\},
\label{vefftot}
\end{align}
\end{widetext}
\noindent
with $\gamma_E\simeq 0.57721$ denoting the Euler-Mascheroni constant. Equation~(\ref{vefftot}), with the boson self-energies given in Eq.~(\ref{Pileading}), provide the necessary tools to explore the effective phase diagram of QCD from the chiral symmetry restoration/breaking point of view.

The matter correction to the boson mass is encoded in the boson self-energy. For a second order (our approach to a crossover due the strict chiral limit) these corrections should cause the thermal boson masses to vanish when the symmetry is restored. This means that in the transition, the effective potential not only develops a minimum but it is also flat (the second derivative vanishes) at $v=0$. This property can be exploited to find the suitable values of the model parameters $a$, $\lambda$ and $g$ at the critical temperature $T_c^0$ for $\mu_B^c=0$. So, the condition to produce a flat effective potential at $T_c$ for $\mu_B=0$ can be written as~\cite{Ayala:2021nhx}
\begin{eqnarray}
6&\lambda&\left(\frac{T_c^2}{12}-\frac{T_c}{4\pi}\left(\Pi_{\text{b}}-a^2\right)^{1/2}\right. \nonumber\\
&+&\left. \frac{a^2}{16\pi^2}\left[\ln\left(\frac{\tilde{\mu}^2}{(4\pi T_c)^2}\right)+2\gamma_E\right]\right)\nonumber\\
&+&g^2T_c^2-a^2=0.
\label{condatTc}
\end{eqnarray}
where, from Eq.~(\ref{Pileading}),
\begin{eqnarray}
\Pi_{\text{b}}(T_c^0,\mu_B^c=0)=\left[\frac{\lambda}{2} + g^2\right](T_c^0)^2.
\label{Pi1formu0}
\end{eqnarray}

We take $\tilde{\mu}=500$ MeV, which is large enough to be considered the largest energy scale in our problem. Notice that the dependence in $\tilde{\mu}$ is only logarithmic, therefore small variations on this parameter do  not change significantly the result. To fix $\lambda$, $g$ and $a$, we look for a set of parameters  such that the solutions of Eq.~(\ref{condatTc}) produce a curve comparable to the LQCD transition curve at $T_c^0\simeq 158$ MeV.  In order to compare the curves, we  use a common parameterization of the LQCD transition curve given by 

\begin{equation}
   \frac{T_c(\mu_B)}{T_c^0}=1-\kappa_2\left(\frac{\mu_B}{T_c^0}\right)^2+\kappa_4\left(\frac{\mu_B}{T_c^0}\right)^4,
    \label{paraLQCD}
\end{equation}
with $\kappa_2=0.0153$ and $\kappa_4=0.00032$~\cite{Borsanyi:2020fev,Guenther:2020jwe}. In other words, fixing the parameters is reduced to finding the set of $\lambda$, $g$ and $a$ such that the coefficients $\kappa_2$ and $\kappa_4$ of their transition curve are comparable to those given by LQCD. Explicitly, we first fix a value of $\lambda$, and find the solution of Eq.~(\ref{condatTc}) for $g$ and $a$ that, from the effective potential in Eq,~(\ref{vefftot}), produce a phase transition at the values of $T_c(\mu_B)$, hereafter simply referred to as $T_c$, and $\mu_B$ given by Eq.~(\ref{paraLQCD}). We then repeat the procedure for other values of $\lambda$.  In this manner, the solutions can be expressed as a relation between the couplings $g$ and $\lambda$ 
\begin{equation}
    g(\lambda)=0.31+1.94 \lambda-2.06 \lambda^2+0.97 \lambda^3-0.20 \lambda^4,
\end{equation}
and a relation between $a$ and $\lambda$
\begin{eqnarray}
 \frac{a(\lambda)}{T_c}&=&0.35+2.08 \lambda-2.21 \lambda^2+1.03 \lambda^3-0.21 \lambda^4.
    \label{arel}
\end{eqnarray}
Since Eq.~(\ref{condatTc}) is non-linear, the set of parameters is not unique.

\begin{figure}[t]
    \centering
    \includegraphics[scale=0.5
    ]{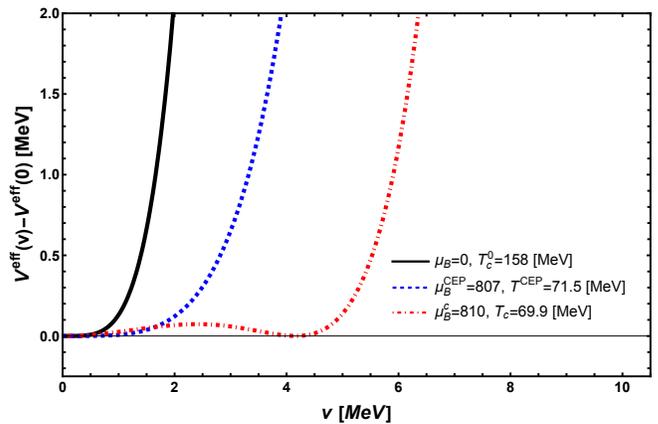}
    \caption{$V^{\text{eff}}$ as a function of the order parameter for different sets of values of $T_c$ and $\mu_B^c$ along the transition curve using $a=148.73$ MeV, $\lambda=1.4$ and $g=0.88$. For $\mu_B=0$ and $T_c^0=158$ MeV the phase transition is second order. For $\mu_B^{\text{CEP}}=807$ MeV and $T^{\text{CEP}}=71.5$ MeV, where the CEP is located, the phase transition becomes first order. 
    }
    \label{Veff}
\end{figure}

\begin{figure}[t]
        \centering
    \includegraphics[scale=0.58]{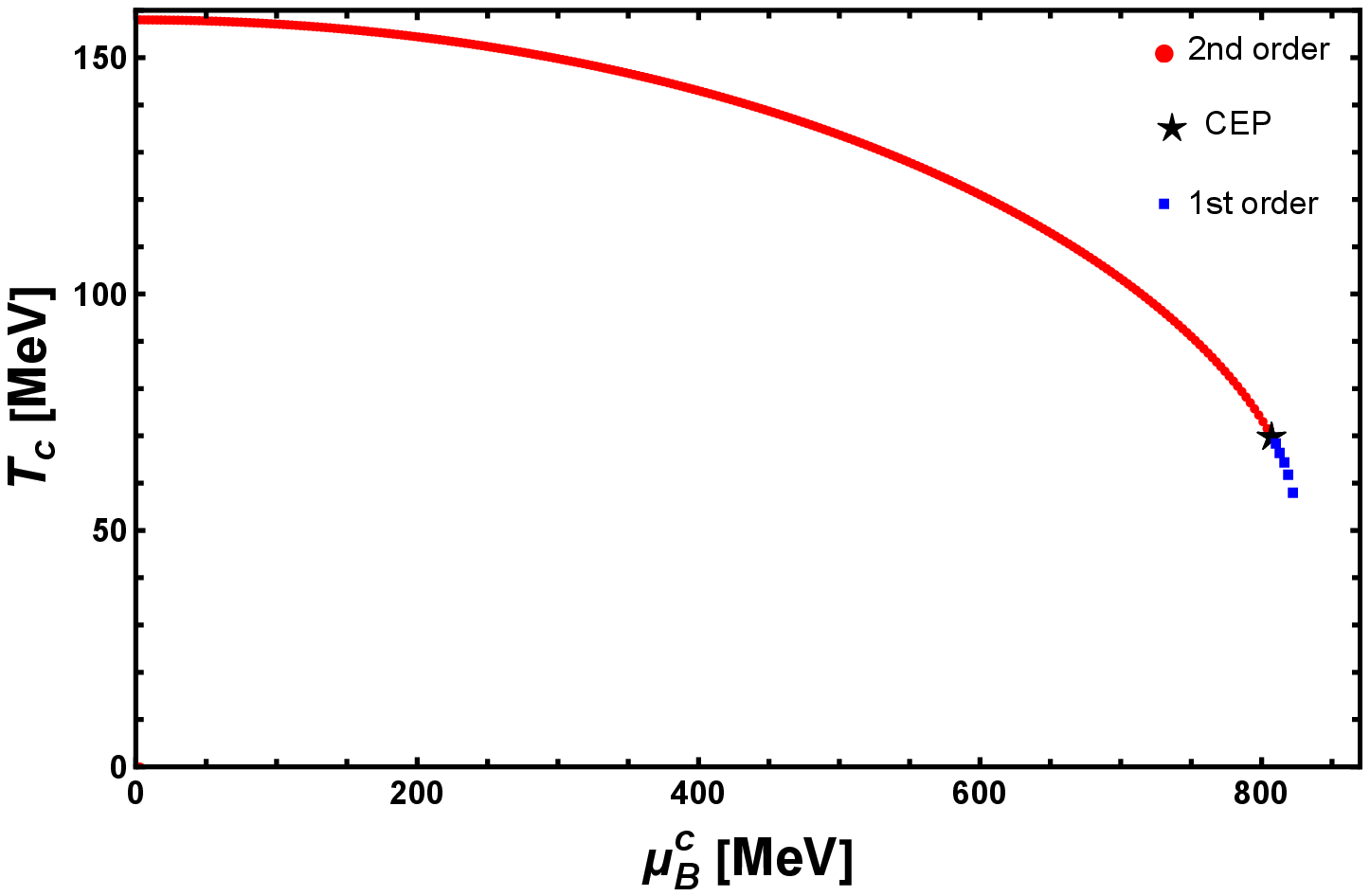}
    \\[\bigskipamount]
        \centering
    \includegraphics[scale=0.58]{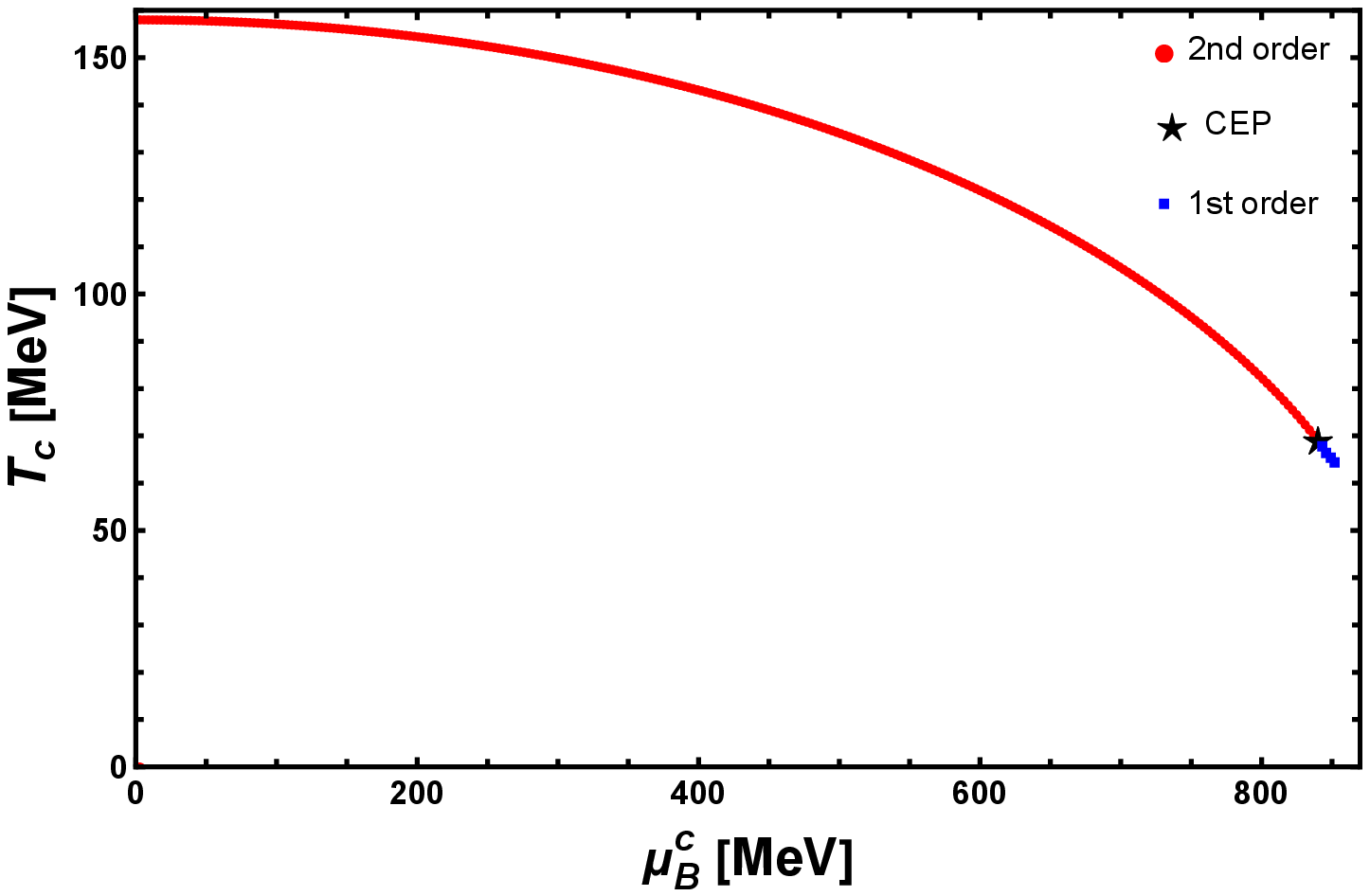}
    \caption{Examples of effective phase diagrams obtained for two choices of the possible sets of parameters $a$, $\lambda$ and $g$. The upper panel is computed with $\lambda=1.4$, $g=0.88$ and $a=148.73$ MeV. The lower panel is computed with $\lambda=0.4$, $g=0.82$ and $a=141.38$ MeV. Notice that the position of the CEP is not significantly altered by varying the choice of parameters. For the allowed range of values of $a$, $\lambda$ and $g$, the CEP location ranges between 786 MeV $<\mu_B^{\text{CEP}}<849$ MeV and 69 MeV $< T^{\text{CEP}}<70.3$ MeV. Adapted from~\cite{Ayala2022}.}
    \label{PD}
\end{figure}

With all the parameters fixed, we can study the properties of the effective potential and find the values of $T_c$ and $\mu_B^c$ where the chiral phase transition takes place. Figure~\ref{Veff} shows the effective potential as a function of the order parameter. We take as examples three different sets of values of $T_c$ and $\mu_B^c$ along the transition curve using $a=148.73$ MeV, $\lambda=1.4$ and $g=0.88$. Notice that for $\mu_B=0$ and $T_c=158$ MeV the phase transition is second order. As $\mu_B$ increases, the phase transition is signaled by a flatter effective potential until the chemical potential and temperature reach values $\mu_B^{\text{CEP}}$ and $T^{\text{CEP}}$, where the effective potential develops a barrier between degenerate minima. For $\mu_B^c>\mu_B^{\text{CEP}}$ and $T_c<T^{\text{CEP}}$, the phase transitions are always first order. This is shown more clearly in Fig.~\ref{PD}, where we show the phase diagrams thus obtained. The upper panel is computed using $\lambda=1.4$, $g=0.88$ and $a=148.73$ MeV. The lower panel is computed with $\lambda=0.4$, $g=0.82$ and $a=141.38$ MeV. The solid (red) lines represent second order phase transitions, our proxy for crossover transitions, whereas the dotted (blue) lines correspond to first order phase transitions. The star symbol represents the location of the CEP. These are the results directly obtained from our analysis. In all cases, we locate the CEP in a region of low temperatures and high quark chemical potential. Also, variations of the model parameters do not change the CEP position appreciably.

In fact, for the allowed range of values of $a$, $\lambda$ and $g$, the CEP location ranges between  786 MeV $<\mu_B^{\text{CEP}}<849$ MeV and 69 MeV $< T^{\text{CEP}}<70.3$ MeV. Now that we have the complete analysis of the phase diagram from the effective potential, in the next section we  introduce the elements necessary to study the phase diagram in terms of baryon number fluctuations.

\section{Baryon number fluctuations}\label{secIII}

\begin{figure}[t]
    \centering
    \includegraphics[scale=0.58
    ]{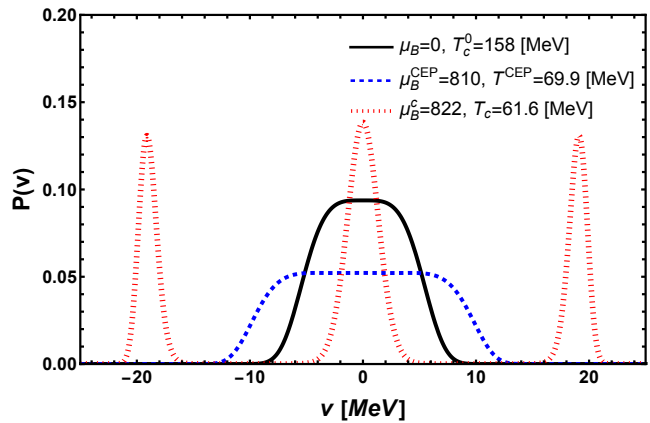}
    \caption{
    Normalized probability distribution for different pairs of $\mu_B^c$, $T_c$ along the transition curve. We use the values of the parameters $\lambda=1.4$, $g=0.88$ and $a=148.73$ MeV. For $\mu_B=0$ and $T_c^0=158$ MeV the probability distribution is Gaussian-like albeit wider. For $\mu_B^{\text{CEP}}$ and $T^{\text{CEP}}$ the probability distribution becomes even wider. For $\mu_B^c>\mu_B^{\text{CEP}}$ and $T_c<T^{\text{CEP}}$, the phase transitions are always first order and the probability distribution develops secondary peaks. Adapted from~\cite{Ayala2022}.
    }
    \label{Boltzmannfactor}
\end{figure}

To study the fluctuations in the number of baryons predicted by the analysis, we start by looking at the probability distribution, which is a function of the order parameter around the equilibrium value in the restored symmetry phase $\langle v\rangle=0$ given by
\begin{eqnarray}
{\mathcal{P}(v)}=\exp\left\{-\Omega V^{\text{eff}}(v)/T\right\},
\label{Boltzfac}
\end{eqnarray}
where, the factor $\Omega$ represents the system volume. For this work, we explored large volumes compared to the typical fireball size created in heavy ion collisions, to simulate the thermodynamic limit. Using $\lambda=1.4$, $g=0.88$ and $a=148.73$ MeV, we illustrate in Fig.~\ref{Boltzmannfactor} the normalized probability distribution for different pairs of $\mu_B^c$, $T_c$ along the transition curve. We have extended the domain of the order parameter making $v\to |v|$ to include negative values and ensure that its average satisfies $\langle v\rangle=0$. Notice that for $\mu_B=0$ and $T_c^0$, for which the phase transition is second order, the probability distribution is Gaussian-like, albeit wider. This is due to the fact that quartic terms in $v$ are important for these values of $\mu_B^c$ and $T_c$ producing a wider distribution around $\langle v\rangle$. As $\mu_B^c$ increases and the phase transition becomes first order for $\mu_B^c>\mu_B^{\text{CEP}}$ and $T_c<T^{\text{CEP}}$, the phase transitions are always first order and the probability distribution develops secondary peaks that reflect the fact that for first order phase transitions the effective potential develops degenerate minima.

\begin{figure}[t]
        \centering
    \includegraphics[scale=0.58]{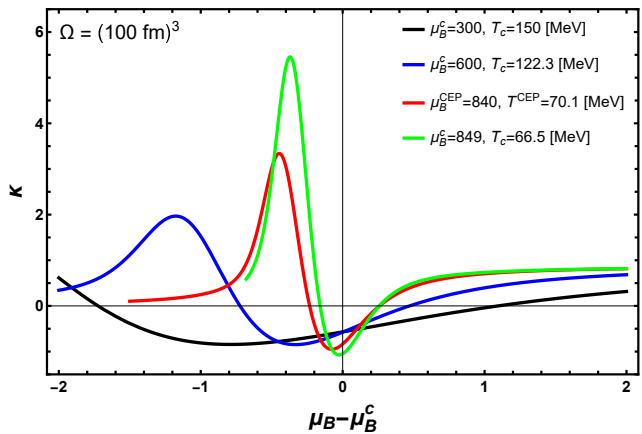}
    \caption{
    Kurtosis  $\kappa$ as functions of $\mu_B-\mu_B^c$ at a fixed value of the corresponding $T_c$ for $\lambda=0.4$, $g=0.82$ and $a=141.38$ MeV and a volume $\Omega=(100\ \text{fm})^3$. For first order phase transitions the Kurtosis show a peaked structure when $\mu_B$ goes across the corresponding critical value of $\mu_B^c$. Adapted from~\cite{Ayala2022}.}
    \label{sandk}
\end{figure}

\begin{figure}[t]
        \centering
    \includegraphics[scale=0.59]{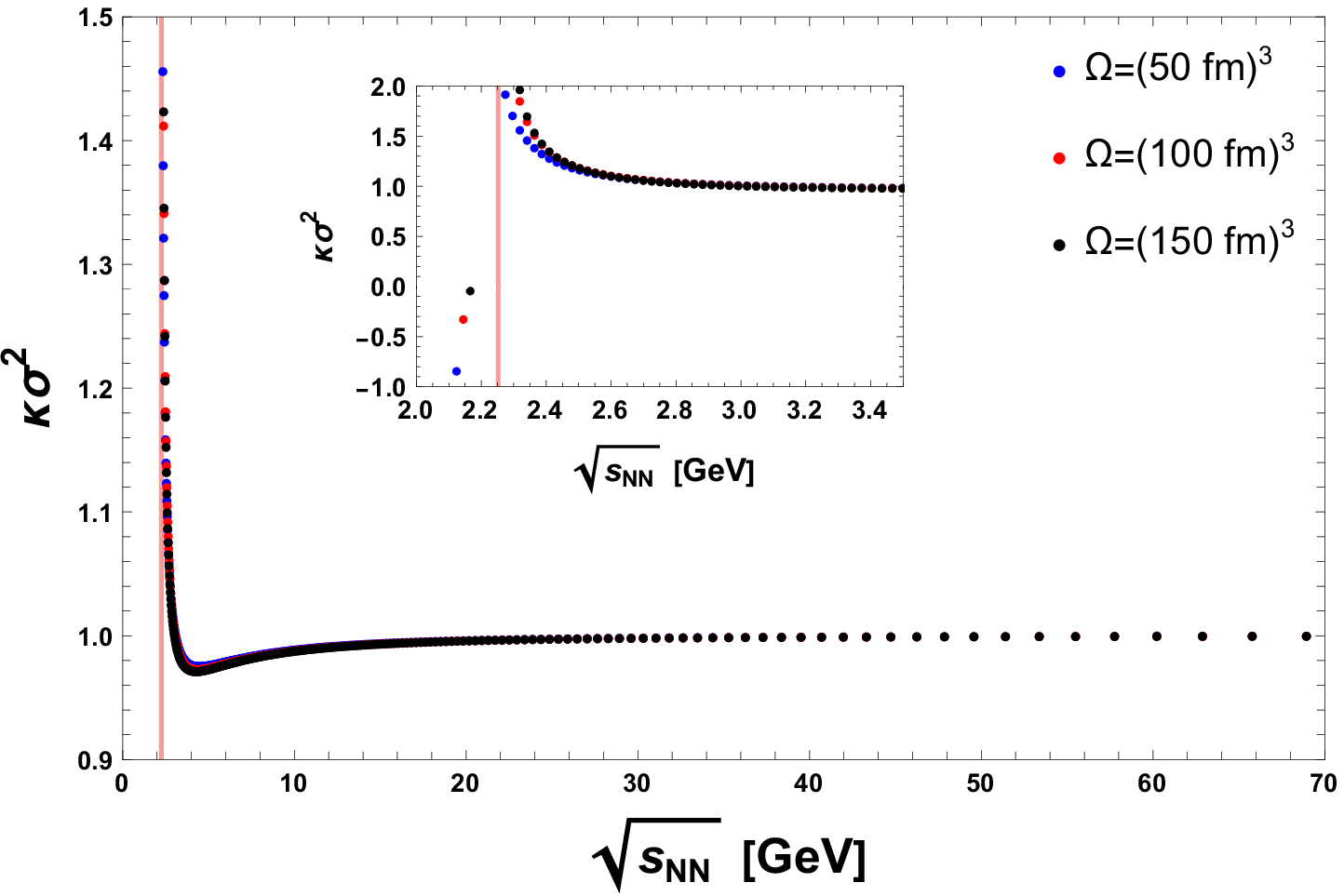}
    \\[\bigskipamount]
        \centering
    \includegraphics[scale=0.59]{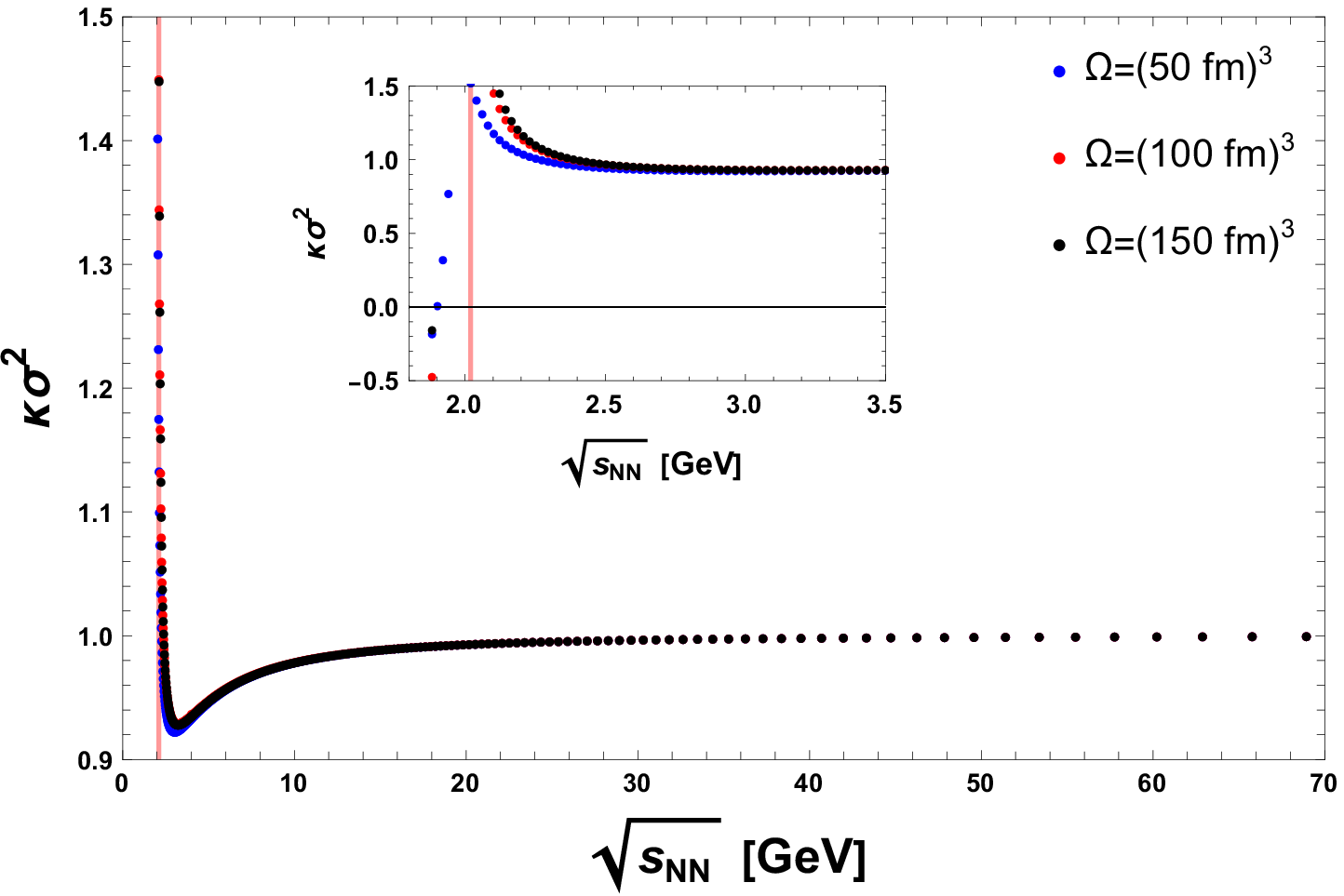}
    \caption{Ratio of the cumulants $C_4/C_2=\kappa\sigma^2$ normalized to the same ratio computed for $\mu_B=0$ and $T=T_c^0$ for three different values of the volume $\Omega$ as a function of the collision energy $\sqrt{s_{NN}}$, using its relation with $\mu_B$ given by Eq.~(\ref{therminatormus}). The upper panel is computed with $a=148.73$ MeV, $\lambda=1.4$ and $g=0.88$. The lower panel is computed with $a=141.38$ MeV, $\lambda=0.4$ and $g=0.82$. In each case the ratio $C_4/C_2$ is independent of $\Omega$ except near the collision energy where we find the CEP, and the high temperature approximation is less accurate. The value of $(\sqrt{s_{NN}})_\text{CEP}\sim 2$ GeV that corresponds to the CEP location for each set of parameters, is represented by the vertical line. The insets show the same ratio of cumulants in a region around $(\sqrt{s_{NN}})_\text{CEP}$. Notice that $\kappa\sigma^2$ significantly drops down as the collision  energy moves from the right to the left across $(\sqrt{s_{NN}})_\text{CEP}$. This behavior is in agreement with recent data~\cite{HADES:2020wpc}. Adapted from~\cite{Ayala2022}.}
    \label{kappasigmaplots}    
\end{figure}

We emphasize that the features of the probability distributions for first order phase transitions are due to the inclusion of the ring diagrams and thus of the plasma screening. If these effects were not included, the development of secondary peaks in the probability distribution would not occur and, therefore, deviations from the Skellam statistic would not be possible. To understand the properties of the probability distribution we calculate the behavior of the fourth order cumulant, also known as kurtosis. Figure~\ref{sandk} shows the kurtosis $\kappa$, as functions of $\mu_B$, for fixed $T$ across the corresponding critical value of $\mu_B^c$, for $\lambda=0.4$, $g=0.82$ and $a=141.38$ MeV. Notice that for second order phase transitions, $\kappa$ is represented by smooth curves. However, when $T$ approaches $T^{\text{CEP}}$, this function shows a peaked structure that becomes more pronounced when the transitions become first order. Notice also that when $T$ goes below $T^{\text{CEP}}$, the kurtosis develops a maximum for $\mu_B>\mu_B^c$. Let us next proceed to describe the behavior of the kurtosis as functions of the collision energy in heavy-ion reactions. To this end, we resort to the relation between the chemical freeze-out value of $\mu_B$ and the collision energy $\sqrt{s_{NN}}$, given by~\cite{Cleymans:2005xv} 
\begin{eqnarray}
   \mu_B(\sqrt{s_{NN}})=\frac{d}{1+e\sqrt{s_{NN}}},
\label{therminatormus}
\end{eqnarray}
where $d=1.308$ GeV and $e=0.273$ GeV$^{-1}$. Figure~\ref{kappasigmaplots} shows the ratio of the cumulants $C_4/C_2=\kappa\sigma^2$, normalized to the same ratio computed for $\mu_B=0$ and $T=T_c$, where $\sigma^2$ is the variance, for three different values of the volume $\Omega$. The upper panel is computed with the set of parameters $a=148.73$ MeV, $\lambda=1.4$ and $g=0.88$, whereas the lower panel corresponds to $a=141.38$ MeV, $\lambda=0.4$ and $g=0.82$. Additionally the value of $\sqrt{s_{NN}}$ for each set of parameters that corresponds to the CEP location is represented by the vertical line.

 Thus, we see that the CEP position is heralded not by the dip of $C_4/C_2$ but for its strong rise as the energy that corresponds to the CEP is approached. A similar result has been found in Ref.~\cite{Mroczek:2020rpm}. Also, as pointed out in Ref.~\cite{Luo:2017uxa}, this non-monotonic behavior cannot be described by many other model calculations~\cite{Xu:2016qjd,He:2016uei}. We emphasize that the reason for this failure is that other models do not include long-range correlations, as we do in this work with the LSMq, which is crucial to describe critical phenomena.
Notice that, as is expected for a ratio of cumulants, in each case the ratio $C_4/C_2$ is independent of the volume $\Omega$ except near the collision energy where we find the CEP, where the high temperature approximation is less accurate.

\begin{figure}[t]
    \centering
    \includegraphics[scale=0.59]{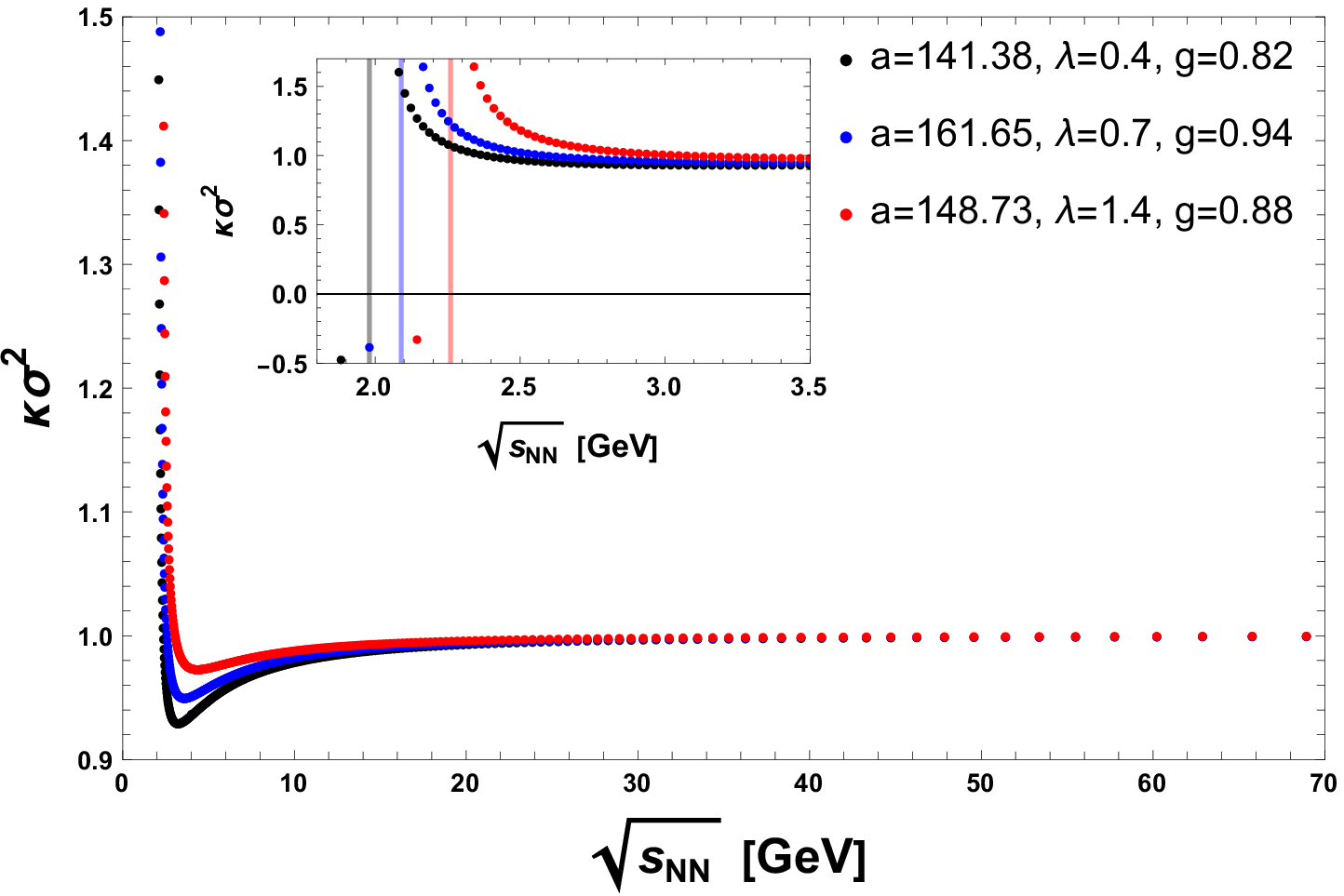}
    \caption{Ratio $C_4/C_2=\kappa\sigma^2$, normalized to the same ratio computed for $\mu_B=0$ and $T=T_c^0$ for $\Omega=(100\ \text{fm})^{3}$ for three different allowed sets of parameters $a$, $\lambda$ and $g$. The dips have different depths. However, the strong rise of the ratio happens for almost the same value of the collision energy $(\sqrt{s_{NN}})_\text{CEP}\sim 2$ GeV. This is shown in the inset where the same ratio of cululants in a region around $(\sqrt{s_{NN}})_\text{CEP}$ is depicted. Notice that $\kappa\sigma^2$ significantly drops down as the collision  energy moves from the right to the left across $(\sqrt{s_{NN}})_\text{CEP}$. This behavior is in agreement with recent HADES data~\cite{HADES:2020wpc}. Adapted from~\cite{Ayala2022}.}
    \label{kslambdas}
\end{figure}
Notice also that the product $\kappa\sigma^2$ significantly drops down for energies lower than the collision energy where the CEP is located, in agreement with recent HADES measurements of net-proton fluctuations at low energies~\cite{HADES:2020wpc}. Finally, Fig.~\ref{kslambdas} shows the ratio $C_4/C_2=\kappa\sigma^2$, normalized to the same ratio computed for $\mu_B=0$ and $T=T_c^0$, for $\Omega=(200\ \text{fm})^{3}$ and for three different allowed sets of parameters $a$, $\lambda$ and $g$.
It is worth mentioning that, even though the dips have different depths, the sharp increase in the ratios occurs for almost the same value of collision energy $\sqrt{s_{NN}}\sim 2$ GeV.

\section{Summary}\label{concl}

In this work we have used the LSMq in the high temperature and chiral limits to explore the QCD phase diagram from the point of view of chiral symmetry restoration. We have computed the finite temperature effective potential up to the contribution of the ring diagrams to account for the plasma screening effects. This model makes an effective description of the system equilibrium distribution that deviates from that of the HRGM, where the ratios of cumulants of even order are always equal to 1. When we include  the plasma screening properties, encoded in the ring diagrams contribution, we find a deviation from the HGR model, since the screening properties describe a key feature of plasma in the transitions, namely, the long-range correlations. We fix the LSMq parameters using conditions at the phase transition for $\mu_B=0$ provided by LQCD calculations, namely the crossover transition temperature $T_c^0$ and the curvature parameters $\kappa_2$ and $\kappa_4$. The phase diagram can be obtained by finding the kind of phase transitions that the effective potential allows when varying $T$ and $\mu_B$. We found that the CEP can be located in the range 786 MeV $<\mu_B^{\text{CEP}}<849$ MeV and 69 MeV $< T^{\text{CEP}}<70.3$ MeV. From the probability distribution obtained using the effective potential, we have computed the behaviour of the kurtosis and found that these cumulants show strong peaks as the CEP is crossed. Finally, we describe the ratio of the cumulants $C_4/C_2=\kappa\sigma^2$ as a function of the collision energy in a heavy-ion collision.

We conclude that the CEP location coincides with a sharp rise in this ratio at $\sqrt{s_{NN}}\sim 2$ GeV. Our studies were performed at the high temperature limit and without including an explicit symmetry breaking term that gives rise to a finite vacuum pion mass. Due to this, the encouraging findings of this work can be extended to provide a more accurate description including an explicit chiral symmetry breaking introduced by a finite pion mass, as well as by relaxing the high-temperature approximation. This is work for the near future that will be reported elsewhere. 
\section*{Acknowledgements}
Support for this work was received in part by UNAM-DGAPA-PAPIIT grant number IG100322 and by Consejo Nacional de Ciencia y Tecnolog\'ia grant numbers A1-S-7655 and A1-S-16215. JJCM acknowledges financial support from the University of Sonora under grant USO315007861. S. H.-O. acknowledges support from the U.S. DOE under Grant No. DE-FG02-00ER41132 and the Simons Foundation under the Multifarious Minds Program Grant No. 557037. METY is grateful for the hospitality of Perimeter Institute where part of this work was carried out and this research was also supported in part by the Simons Foundation through the Simons Foundation Emmy Noether Fellows Program at Perimeter Institute. Research at Perimeter Institute is supported in part by the Government of Canada and by the Province of Ontario.

%
\nocite{*}
\bibliography{ref}

\end{document}